\documentstyle[aps,twocolumn,psfig,pra]{revtex}

\newcommand{\ket}[1]{|{#1}\rangle}

\begin{document}

\draft

\title{ Spin squeezing as a measure of entanglement in a two qubit 
system}
\author{A. Messikh$^{1}$, Z. Ficek$^{1,2}$ and M.R.B. Wahiddin$^{1}$}
\address{$^{1}$ Centre for Computational and Theoretical Sciences,
Kulliyyah of Science, International Islamic University Malaysia,
53100 Kuala Lumpur, Malaysia
\\ [8pt]
$^{2}$ Department of Physics, The University of Queensland, Brisbane, 
QLD 4072, Australia}

\date{\today}

\maketitle

\begin{abstract}
    We show that two definitions of spin squeezing extensively used in the 
    literature [M. Kitagawa and M. Ueda, Phys. Rev. A {\bf 47}, 5138 
    (1993) and D.J. Wineland {\it et al.}, Phys. Rev. A {\bf 50}, 67 
    (1994)] give different predictions of entanglement in the two-atom 
    Dicke system. We analyze differences between the definitions 
    and show that the Kitagawa and Ueda's spin squeezing parameter is a 
    better measure of entanglement than the commonly used spectroscopic 
    spin squeezing parameter. We illustrate this relation by examining 
    different examples of a driven two-atom Dicke system 
    in which spin squeezing and entanglement arise dynamically. 
    We give an explanation of the source of the difference 
    in the prediction of entanglement using the negativity criterion for
    entanglement. For the examples discussed, we find that  
    the Kitagawa and Ueda's spin squeezing parameter is the sufficient and
    necessary condition for entanglement.
\end{abstract}




Spin squeezing results from quantum correlations 
between atomic spins and have received a great deal of attention in recent
years~\cite{ku,wbih,sm,uok,wm,tmw,jk,ws,dws}. The interest in spin squeezing 
arises not only from the fact that it exhibits
reduced fluctuations of the collection of atomic spins below the fundamental 
spin noise limit, but also from the possibility of interesting novel 
applications in interferometry, high precision spectroscopy and atomic 
clocks. Recently, S\o rensen {\it et al.}~\cite{sdcz} have proposed spin 
squeezing as a measure of entanglement in multi-atom systems, which 
opens further applications in the area of quantum information and quantum 
computation~\cite{nc}. The advantage of spin squeezing over 
the well known entanglement measures, such as concurrence~\cite{woo}
and negativity~\cite{per,horo} is that spin squeezing can be used as a 
measure of entanglement in multi-atom systems, whereas the former measures
can be applied only to two particle (two qubit) systems.
Hald {\it et al.}~\cite{hal} recently 
reported preparation of an entangled multiatom state via quantum 
state transfer from squeezed light to a collection of atomic spins. 
Kuzmich {\it et al.}~\cite{kbm} have proposed a scheme to produce 
spin squeezed states via a quantum nondemolution 
measurement technique and spin noise reduction using this method has 
been experimentally observed~\cite{kmb}.

There are, however, two different definitions of the spin squeezing 
parameter frequently used in the literature; the Kitagawa and Ueda's 
spin squeezing parameter defined as~\cite{ku}
\begin{eqnarray}
     \xi_{n_{i}}^{S} =\frac{2}{S}\langle \left(\Delta
     S_{\vec{n}_{i}}\right)^{2}\rangle_{\perp} ,\qquad i=1,2,\label{eq1}
\end{eqnarray}
and the spectroscopic spin squeezing parameter introduced in the 
context of Ramsey spectroscopy as~\cite{wbih}
\begin{eqnarray}
     \xi_{n_{i}}^{R} =\frac{2S\langle \left(\Delta
     S_{\vec{n}_{i}}\right)^{2}\rangle_{\perp}}
     {\langle S_{\vec{n}_{3}}\rangle^{2}} ,\label{eq2}
\end{eqnarray}
where $S$ is the total spin of the system, $\vec{n}_{1}, \vec{n}_{2}$
and $\vec{n}_{3}$ are
three mutually orthogonal unit vectors oriented such that the mean
value of one of the spin components, assumed here 
$\langle S_{\vec{n}_{3}}\rangle$, is different from zero, while the 
other components $S_{\vec{n}_{1}}$ and
$S_{\vec{n}_{2}}$ have zero mean values. The variance
$\langle \left(\Delta S_{\vec{n}_{i}}\right)^{2}\rangle_{\perp}$ is 
calculated in the plane orthogonal to the mean spin direction.
A multiatom system in a coherent state has variances normal to the mean
spin direction equal to the standard quantum limit of $S/2$. In this
case, $\xi_{\vec{n}_{i}}^{S}=1$. A system with the variance reduced below
the standard quantum limit is characterized by $\xi_{\vec{n}_{i}}^{S}<1$, 
that is spin squeezed in a direction normal to the mean spin direction.
With the parameter (\ref{eq2}), spin squeezing is manifested by 
$\xi_{\vec{n}_{i}}^{R}<1$ which indicates a reduction in the 
frequency noise.
Since the mean value $|\langle  S_{\vec{n}_{3}}\rangle | \leq S$, 
it follows that the parameters (\ref{eq1}) and (\ref{eq2}) do not describe
the same spin squeezing, that a spectroscopic spin squeezing 
$\xi_{\vec{n}_{i}}^{R} <1$ implies $\xi_{\vec{n}_{i}}^{S} <1$, but not 
vice versa. It should be noted here that in general the spin squeezing 
parameters (\ref{eq1}) and (\ref{eq2}) are sufficient but not necessary 
conditions for entanglement, that one can create entanglement without 
spin squeezing~\cite{ban,zsl,mfw}.

In studying the relation between entanglement and spin squeezing, we 
discovered that the two definitions of spin squeezing give somewhat 
different predictions of entanglement in the two-atom Dicke system.
It is the purpose of this Brief Report to point out that the Kitagawa 
and Ueda's spin squeezing parameter (\ref{eq1}) is a better measure of 
entanglement than the the spectroscopic spin squeezing parameter 
(\ref{eq2}). Specifically, we will show that there is a large class of 
processes for which the Kitagawa and Ueda's parameter is 
the sufficient and necessary condition for entanglement.
It was quite surprising to find this connection, since the 
spectroscopic spin squeezing parameter (\ref{eq2}) is commonly used in 
the literature to compute spin squeezing and 
entanglement in multi-atom systems. The spin squeezing is currently 
the widely accepted measure of multi-atom entanglement, so we believe 
that a detailed analysis of the relation between entanglement and these 
two definitions of spin squeezing is of general interest.

We illustrate our considerations of the relation between entanglement and 
the spin squeezing parameters in a simple model of the two-atom (two 
qubit) Dicke system which consists of two identical 
atoms confined to a volume with dimensions much smaller than the 
wavelength of the atomic transitions~\cite{dic,leh,ft}. In this limit, 
the atomic dipole moments evolve on the time scale much shorter than any 
$\vec{S}^{2}$-breaking relaxation mechanism, so that the total spin 
$\vec{S}^{2}$ of the system is conserved during the evolution. 
Each atom is assumed to have only two 
energy levels, ground level $|g_{i}\rangle$ and excited level 
$|e_{i}\rangle$ $(i=1,2)$, separated by an energy $\hbar \omega_{0} 
\equiv E_{e}-E_{g}$, where $\omega_{0}$ is the atomic transition 
frequency. The atomic levels $|g_{i}\rangle$ and $|e_{i}\rangle$ are 
eigenstates of the energy operator $S_{i}^{z}$ with eigenvalues $-1/2$ 
and $1/2$, respectively. Although the two-atom Dicke system is admittedly 
an elementary model, it offers an insight into the fundamentals 
of spin squeezing generation process and entanglement creation in 
multiatom systems. Because of its simplicity, we will obtain  
exact solutions for the density matrix elements of the system, which 
will allow us to study in details the relation between the atomic 
coherence properties and the amount of spin squeezing.

In the absence of external driving fields, the two-atom Dicke 
system is equivalent to a cascade multilevel system composed of 
three energy levels~\cite{dic,leh}
\begin{eqnarray}
\ket g &=& \ket {g_{1}}\ket {g_{2}} ,\nonumber \\
\ket s &=& \left( \ket {e_{1}}\ket {g_{2}}
+\ket {g_{1}}\ket {e_{2}}\right)/\sqrt{2} ,\nonumber \\
\ket e &=& \ket {e_{1}}\ket {e_{2}} ,\label{eq3}
\end{eqnarray}
with energies $E_{g}=-\hbar \omega_{0}, E_{s}= 0$ and $E_{e}=\hbar 
\omega_{0}$, respectively. The energy levels (\ref{eq3}) are known as 
the Dicke collective states~\cite{dic}. The ground $\ket g$ and the 
upper $\ket e$ 
levels are product states of the individual atoms, whereas the 
intermediate state $\ket s$ is a maximally entangled state
of the two-atom system. The state is a linear superposition of the
product states which cannot be separated into product states of
the individual atoms.

In our analysis, we assume that the atoms are driven by two fields
of fixed phases but different statistics. 
These are a coherent laser field of the (real) 
Rabi frequency $\Omega$, and a broadband squeezed vacuum field which 
is known to produce strong two-photon coherences in atomic systems. 
To keep the mathematical complications to a minimum, we assume that 
the angular frequency $\omega_{L}$ 
of the laser field and the carrier frequency $\omega_{s}$ of the 
squeezed field are equal to the atomic transition frequency, i.e. the 
detunings $\Delta_{L} =\omega_{L}-\omega_{0}$ and 
$\Delta_{s}=\omega_{s}-\omega_{0}$ are zero. We will examine the 
relation between entanglement and the spin squeezing parameters in 
three different models of the interaction in which entanglement and 
spin squeezing arise dynamically. In the first, the atoms interact only
with the squeezed field $(\Omega =0)$. In this case, one achieves spin 
squeezing and entanglement induced by the two-photon coherences with 
no contribution from the one-photon coherences. In the second model, 
the atoms interact only with the coherent field. Here, both the one 
and two-photon coherences contribute to the spin squeezing and 
entanglement. In the third model, the atoms interact simultaneously 
with both fields. In this case the squeezed field enhances the two-photon 
coherences.

To calculate the variances and the mean values of the spin components 
appearing in Eqs.~(\ref{eq1}) and (\ref{eq2}), we apply the master equation 
of the driven two-atom Dicke system, which in the interaction picture 
is given by~\cite{dfs}
\begin{eqnarray}
       \frac{\partial \hat{\rho}}{\partial t} &=& \frac{i}{\hbar}
       \left[H_{s}, \hat{\rho}\right] \nonumber \\
       &-& \frac{1}{2}\Gamma
       \left(N+1\right)\left(S^{+}S^{-}\hat{\rho} +\hat{\rho}S^{+}S^{-}
       -2S^{-}\hat{\rho}S^{+}\right) \nonumber \\
       &-& \frac{1}{2}\Gamma
       N\left(S^{-}S^{+}\hat{\rho} +\hat{\rho}S^{-}S^{+}
       -2S^{+}\hat{\rho}S^{}\right) \nonumber \\
       &+& \frac{1}{2}\Gamma
       M\left(S^{+}S^{+}\hat{\rho} +\hat{\rho}S^{+}S^{+}
       -2S^{+}\hat{\rho}S^{+}\right) \nonumber \\
       &+& \frac{1}{2}\Gamma
       M^{\ast}\left(S^{-}S^{-}\hat{\rho} +\hat{\rho}S^{-}S^{-}
       -2S^{-}\hat{\rho}S^{-}\right) ,\label{eq4}
\end{eqnarray}
where $\Gamma$ is the spontaneous emission rate of the atoms, 
$S^{\pm}=S^{\pm}_{1}+S^{\pm}_{2}$ are the collective atomic spin 
operators, and $H_{s}=-i\hbar (\Omega/2) \left(S^{+}-S^{-}\right)$ is 
the interaction Hamiltonian between the atoms and the laser field. 
The parameters $N$ and $M$ characterize the squeezed vacuum field,
such that $N$ is the number of photons in the squeezed
modes, $M=|M|{\rm exp}(i\phi_{s})$ is the magnitude of
two-photon correlations between the vacuum modes, and
$\phi_{s}$ is the phase of the squeezed field. For simplicity, we 
set the squeezing phase $\phi_{s}=0$ (or $\pi$) so that the squeezing 
parameter $M$ is real.

In order to analyze the relation between entanglement and the spin 
squeezing parameters, we express the parameters (\ref{eq1}) and 
(\ref{eq2}) in terms of the density matrix elements of the system. 
Since the driving fields are on resonance with the atomic transition 
and $M^{\ast}=M$, the stationary off-diagonal density matrix 
elements (coherences) are real, or equivalently, the Bloch 
vector has the components 
$\vec{B} =(\langle S_{x}\rangle ,0,\langle S_{z}\rangle )$, 
where $S_{x}=(S^{+}+S^{-})/2$ and $S_{z}=S^{z}_{1} +S^{z}_{2}$. Thus, 
we can study spin squeezing by a single rotation of the nonzero spin 
components around the $y$-axis. Let $\vec{n}_{3}$ be the direction of 
the total spin in the new (rotated) reference frame. Then the 
variances calculated in the directions $\vec{n}_{1}$ and 
$\vec{n}_{2}$ perpendicular to the direction of the total spin can be 
written as
\begin{eqnarray}
    \langle \left(\Delta S_{\vec{n}_{1}}\right)^{2}\rangle_{\perp} 
    &=& \langle S_{z}^{2}\rangle \sin^{2}\alpha +\langle 
     S_{x}^{2}\rangle \cos^{2}\alpha -\langle S_{x}S_{z}\rangle \sin 
     2\alpha ,\nonumber \\
    \langle \left(\Delta 
    S_{\vec{n}_{2}}\right)^{2}\rangle_{\perp} &=& \langle 
    S_{y}^{2}\rangle ,\label{eq5}
\end{eqnarray}
where $\tan \alpha = \langle S_{x}\rangle/\langle S_{z}\rangle$.

A simple calculation using Eqs.~(\ref{eq1}), (\ref{eq2}) and 
(\ref{eq5}) shows that the Kitagawa and Ueda's parameter becomes
\begin{eqnarray}
    \xi_{\vec{n}_{1}}^{S} &=& 
    2\left(1-\rho_{ss}\right)\sin^{2}\alpha +\left(1+\rho_{ss} 
    +2\rho_{eg}\right)\cos^{2}\alpha ,\nonumber \\
    \xi_{\vec{n}_{2}}^{S} &=& 1+\rho_{ss} -2\rho_{eg} 
      ,\label{eq6}
\end{eqnarray}
whereas for the spectroscopic spin squeezing parameter takes the form 
\begin{eqnarray}
    \xi_{\vec{n}_{1}}^{R} &=& \left[2\left(1-\rho_{ss}\right)\sin^{2}\alpha 
    +\left(1+\rho_{ss} +2\rho_{eg}\right)\cos^{2}\alpha\right]/U^{2} 
    ,\nonumber \\
    \xi_{\vec{n}_{2}}^{R} &=& \left(1+\rho_{ss} -2\rho_{eg}\right)/U^{2}  
    ,\label{eq7}
\end{eqnarray}
where $U=(\rho_{ee}-\rho_{gg})\cos\alpha 
+2^{-1/2}(\rho_{es}+\rho_{sg} +\rho_{se} +\rho_{gs})\sin\alpha$.

From the structure of Eqs.~(\ref{eq6}) and (\ref{eq7}) it is clear
that the necessary 
condition to obtain spin squeezing is to create two-photon 
coherences $\rho_{ge}$. For $\rho_{eg}<0$, the right-hand sides of 
$\xi_{\vec{n}_{1}}^{S}$ and $\xi_{\vec{n}_{1}}^{R}$ can be  
less than 1, whereas the right-hand sides of $\xi_{\vec{n}_{2}}^{S}$ 
and $\xi_{\vec{n}_{2}}^{R}$ are always greater than 1. Thus, spin 
squeezing can be observed only in the $\xi_{\vec{n}_{1}}^{S}$ and 
$\xi_{\vec{n}_{1}}^{R}$ components. On the other hand, for 
$\rho_{eg}>0$, the right-hand sides of only $\xi_{\vec{n}_{2}}^{S}$ 
and $\xi_{\vec{n}_{2}}^{R}$ can be less than 1.

Having introduced the spin squeezing parameters in terms of the 
density matrix elements, we now turn to our central problem to 
determine which of the spin squeezing parameters is a better measure 
of entanglement and what is the degree of entanglement.
Consider first the two-atom Dicke system driven by a 
broadband squeezed vacuum field alone $(\Omega =0)$. In this case, 
the master equation 
(\ref{eq4}) leads to the following nonzero steady-state solutions for 
the density matrix elements~\cite{ft}
\begin{eqnarray}
\rho_{ee} &=& \frac{N^{2}\left(2N+1\right)
-\left(2N-1\right)|M|^{2}}{\left(2N+1\right)
\left(3N^{2} +3N +1 -3|M|^{2}\right)} ,\nonumber \\
\rho_{ss} &=& \frac{N\left(N+1\right)
-|M|^{2}}{3N^{2} +3N +1 -3|M|^{2}} ,\nonumber \\
\rho_{eg} =\rho_{ge} &=& \frac{|M|}{\left(2N+1\right)
\left(3N^{2} +3N +1 -3|M|^{2}\right)} .\label{eq8}
\end{eqnarray}
This equation expresses the steady-state of the system in terms of 
the intensity and the two-photon correlations characteristic of a 
squeezed field. Since the one-photon coherences are zero, we can 
easily verify that $\langle S_{z}\rangle \neq 0$ and 
$\langle S_{x}\rangle =\langle S_{y}\rangle =0$. This implies that 
we can determine spin squeezing in the $xy$-plane without any rotation
$(\alpha =0)$. In this case $(n_{1},n_{2},n_{3})=(x,y,z)$.

With the steady state solution (\ref{eq8}), the density matrix of 
the system in the basis of the product states 
$\{|e_{1},e_{2}\rangle, |e_{1},g_{2}\rangle, |g_{1},e_{2}\rangle,
|g_{1},g_{2}\rangle \}$ takes the form
\begin{eqnarray}
\hat{\rho} &=& \left(
\begin{array}{cccc}
\rho_{ee} & 0 & 0 & \rho_{eg} \\
0 & \frac{1}{2}\rho_{ss} & \frac{1}{2}\rho_{ss} & 0 \\
0 & \frac{1}{2}\rho_{ss} & \frac{1}{2}\rho_{ss} & 0 \\
\rho_{ge} & 0 & 0 & \rho_{gg}
\end{array}
\right) , \label{eq9}
\end{eqnarray}
where $\rho_{gg} =1-\rho_{ee}-\rho_{ss}$.

Given the density matrix, it is possible to calculate the entanglement 
between the atoms. To quantify the degree of entanglement, we use
the negativity criterion for entanglement~\cite{per,horo} and 
find that the eigenvalues of the partial transposition of $\hat{\rho}$
are
\begin{eqnarray}
     \lambda_{1\pm} &=& \frac{1}{2}\rho_{ss} \pm |\rho_{eg}| ,\nonumber
     \\
     \lambda_{2\pm} &=& \frac{1}{2}\{\left(\rho_{ee}+\rho_{gg}\right) \pm
     \left[\left(\rho_{ee}-\rho_{gg}\right)^{2}
     +\rho_{ss}^{2}\right]^{\frac{1}{2}}\} .\label{eq10}
\end{eqnarray}
From this it readily follows that $\lambda_{1+}$ and $\lambda_{2+}$ are 
always positive. Moreover, it is easily verified that with the solution 
(\ref{eq8}), the eigenvalue $\lambda_{2-}$ is positive for all values of 
the parameters involved. Thus, the system exhibits entanglement when 
$|\rho_{eg}| >\rho_{ss}/2$, and then the degree of entanglement is
\begin{eqnarray}
     E = {\rm max}\left(0, -2\lambda_{1-}\right) 
     =2|\rho_{eg}| -\rho_{ss} . \label{eq11}
\end{eqnarray}
It is evident by comparison of Eq.~(\ref{eq11}) with Eqs. (\ref{eq6}) 
and (\ref{eq7}) that the condition for entanglement $(E>0)$ is completely 
equivalent to the condition for spin squeezing predicted by 
the Kitagawa and Ueda's parameter, and there is a simple relation
\begin{eqnarray}
    E = 1 -\xi_{\vec{n}_{2}}^{S} .\label{eq12}
\end{eqnarray}
A value of $\xi_{\vec{n}_{2}}^{S}<1$ indicates spin squeezing and at 
the same moment there is entanglement $(E>0)$ between the atoms.
In addition, the degree of entanglement is equal to the degree of
spin squeezing. Thus, we conclude that the Kitagawa and Ueda's parameter 
is the sufficient and necessary condition for entanglement induced by a 
squeezed vacuum field. 

\begin{figure}[h]
\begin{center}
\mbox{ \psfig{file=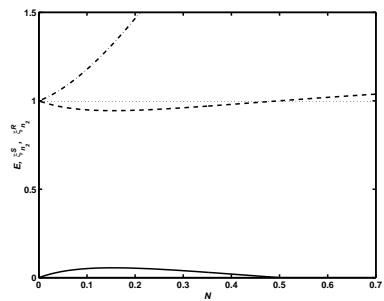,width=5cm}}
\end{center}
\caption{Entanglement measure $E$ (solid line) and the spin
squeezing parameters $\xi_{\vec{n}_{2}}^{S}$ (dashed line) and 
$\xi_{\vec{n}_{2}}^{R}$ (dashed-dotted line) as a function of $N$ 
for the classical squeezed field with $M=N$. }
\label{fig1}
\end{figure}

\begin{figure}[h]
\begin{center}
\mbox{ \psfig{file=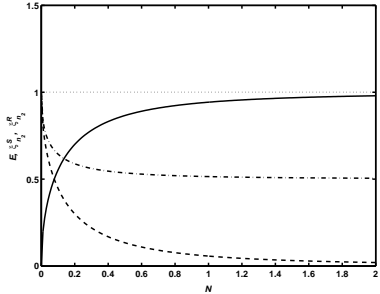,width=5cm}}
\end{center}
\caption{Entanglement measure $E$ (solid line) and the spin
squeezing parameters $\xi_{\vec{n}_{2}}^{S}$ (dashed line),  
$\xi_{\vec{n}_{2}}^{R}$ (dashed-dotted line) as a function of $N$ for 
the quantum squeezed field with $M=\sqrt{N(N+1)}$. }
\label{fig2}
\end{figure}

The above considerations are illustrated in Figs.~\ref{fig1} and 
\ref{fig2}, which show the entanglement measure $E$ and the 
parameters $\xi_{\vec{n}_{2}}^{S}$ and $\xi_{\vec{n}_{2}}^{R}$ for two 
different types of the squeezed field. In Fig.~\ref{fig1}, we plot the 
entanglement measure and the spin squeezing parameters for a 
classical squeezed field with the correlations $M=N$. The classical 
squeezed field is characterized by an anisotropic distribution of 
noise with the noise reduced in some directions, but not below the 
standard vacuum level. The figure shows that 
$\xi_{\vec{n}_{2}}^{R}>1$ for all $N$, but $\xi_{\vec{n}_{2}}^{S}$ is 
less than 1 for $N<1/2$, and also an entanglement appears in the same 
range of $N$. This shows that $\xi_{\vec{n}_{2}}^{S}$ correctly predicts 
entanglement, while with the parameter $\xi_{\vec{n}_{2}}^{R}$ one 
could observe entanglement without spin squeezing.

In Fig.~\ref{fig2}, we plot $E$ and the spin squeezing parameters for 
a quantum squeezed field with perfect correlations $M^{2}=N(N+1)$. 
Since $\rho_{ss}=0$ and the inequality $|\rho_{eg}|>0$ always holds, 
both parameters $\xi_{\vec{n}_{2}}^{S}$ and $\xi_{\vec{n}_{2}}^{R}$ 
are less than 1 for the entire range of $N$. Thus, both parameters 
predict entanglement and spin squeezing for all $N$. However, the 
degree of entanglement is equal to the degree of spin squeezing 
given by $\xi_{\vec{n}_{2}}^{S}$. The entanglement and spin squeezing 
increase with increasing $N$ and attain their maximal values $E=1$ and
$\xi_{\vec{n}_{2}}^{S}=0$ for large $N$. 

It is easy to show that the entanglement created by the quantum 
squeezed field is related to the pure two-atom squeezed 
state. Under the squeezed field excitation, there
are entangled states generated which can be found by the
diagonalization of the density matrix (\ref{eq9}). We find that the 
two-photon coherences produce entangled states
\begin{eqnarray}
|\Psi_{+}\rangle &=& \left[\left(\Pi_{+}-\rho_{ee}\right)|g\rangle
+\rho_{eg}|e\rangle
\right]/{\cal{N}}_{+} ,\nonumber \\
|\Psi_{-}\rangle &=& \left[\rho_{ge}|g\rangle +
\left(\Pi_{-}-\rho_{gg}\right)|e\rangle
\right]/{\cal{N}}_{-} ,\label{eq13}
\end{eqnarray}
where ${\cal{N}}_{\pm}$ are the normalization constants, and
\begin{eqnarray}
\Pi_{\pm} &=& \frac{1}{2}\left(\rho_{gg}+\rho_{ee}\right)
\pm\frac{1}{2}\left[\left(\rho_{gg}-\rho_{ee}\right)^{2}
+4\left|\rho_{eg}\right|^{2}\right]^{\frac{1}{2}} \label{eq14}
\end{eqnarray}
are the populations of the entangled states.

It is evident from (\ref{eq13}), that the two-photon coherences create
entangled states which are linear superpositions of the
ground state $|g\rangle$ and the upper state $|e\rangle$.
Thus, entanglement created by a squeezed field is associated with the 
two-photon entangled states $|\Psi_{\pm}\rangle$. Note that the 
steady-state with the classical squeezed field is a mixed state with 
the populations $\rho_{ss}\neq 0$ and $\Pi_{\pm}\neq 0$, whereas for the 
quantum squeezed field $\rho_{ss}=0$, $\Pi_{-}=0$, and then 
the stationary state of the system is a pure state~\cite{pk90,pa}
\begin{eqnarray}
|\Psi_{+}\rangle = \frac{1}{\sqrt{2N+1}}
\left[\sqrt{N+1}|g\rangle
+\sqrt{N}|e\rangle\right] .\label{eq15}
\end{eqnarray}
The pure state is a non-maximally entangled state, and reduces to a
maximally entangled state for $N\gg 1$. According to Fig.~\ref{fig2}, 
the pure state admits of the largest amount of 
entanglement and for $N\gg 1$, the optimum entanglement $E=1$ can be 
achieved. In Fig.~\ref{fig1}, the mixed state admits of lower level of 
entanglement, and also the entanglement occurs in a restricted range 
of the intensity~$N$.

We now consider the second model in which the system is driven 
by the coherent field $(\Omega \neq 0)$ in the absence of the squeezed 
field $(N=M=0)$. This is an interesting example where one can create 
spin squeezing and entanglement with the linear Hamiltonian $H_{s}$.
Typical schemes considered for the generation of spin squeezing involve
quadratic Hamiltonians~\cite{ku,wbih,sm,uok,wm,tmw,jk,ws,dws,sdcz}.
After straightforward but lengthy calculations, we 
find the following steady-state solutions for the density matrix 
elements
\begin{eqnarray}
\rho_{ee} &=& \Omega^{4}/D ,\nonumber \\
\rho_{ss} &=&  \left(\Omega^{4} +2\Gamma^{2}\Omega^{2}\right)/D 
,\nonumber \\
\rho_{sg} =\rho_{gs} &=&  \sqrt{2}\Gamma\Omega 
\left(\Omega^{2}+2\Gamma^{2}\right)/D ,\nonumber \\
\rho_{es} =\rho_{se} &=&  \sqrt{2}\Gamma \Omega^{3}/D ,\nonumber \\
\rho_{eg} =\rho_{ge} &=& 2\Gamma^{2}\Omega^{2}/D ,\label{eq16}
\end{eqnarray}
where $D= 3\Omega^{4}+4\Gamma^{2}\Omega^{2}+4\Gamma^{4}$.

In order to analyze the relationship between entanglement and spin 
squeezing parameters, we write the density matrix of the system
\begin{eqnarray}
\hat{\rho} &=& \left(
\begin{array}{cccc}
\rho_{ee} & \frac{1}{\sqrt{2}}\rho_{es} & \frac{1}{\sqrt{2}}\rho_{es}
& \rho_{eg} \\
\frac{1}{\sqrt{2}}\rho_{se} & \frac{1}{2}\rho_{ss}
& \frac{1}{2}\rho_{ss} & \frac{1}{\sqrt{2}}\rho_{sg} \\
\frac{1}{\sqrt{2}}\rho_{se} & \frac{1}{2}\rho_{ss} & 
\frac{1}{2}\rho_{ss} & \frac{1}{\sqrt{2}}\rho_{sg} \\
\rho_{ge} & \frac{1}{\sqrt{2}}\rho_{gs} & \frac{1}{\sqrt{2}}\rho_{gs} 
& \rho_{gg}
\end{array}
\right) , \label{eq17}
\end{eqnarray}
and again make use of the negativity criterion for entanglement. There 
are obviously four eigenvalues of the partial transposition of the 
matrix (\ref{eq17}). It is straightforward to show that one of the 
eigenvalues is 
\begin{eqnarray}
     p_{1} &=& \frac{1}{2}\rho_{ss} - |\rho_{eg}| ,\label{eq18}
\end{eqnarray}
whereas the remaining eigenvalues are the three roots of the cubic 
equation
\begin{eqnarray}
   && p^{3} -(1-\frac{1}{2}\rho_{ss}+\rho_{eg})p^{2} 
   \nonumber \\
   && +[\left(1-\rho_{ss}\right)(\frac{1}{2}\rho_{ss}
   +\rho_{eg}) +\rho_{ee}\rho_{gg} -\frac{1}{4}\rho_{ss}^{2} 
   -\rho_{es}^{2}-\rho_{sg}^{2}]p \nonumber \\
   && - (\frac{1}{2}\rho_{ss} +\rho_{eg}) 
   (\rho_{ee}\rho_{gg}-\frac{1}{4}\rho_{ss}^{2}) 
   +\rho_{gg}\rho_{es}^{2} +\rho_{ee}\rho_{sg}^{2} \nonumber \\ 
   && -\rho_{ss}\rho_{sg}\rho_{es} = 0 .\label{eq19}
\end{eqnarray}
It is easily verified from Eqs.~(\ref{eq16}) and (\ref{eq19}) that 
the roots $p_{i}$ are real and positive for all values of $\Omega$.

\begin{figure}[h]
\begin{center}
\mbox{ \psfig{file=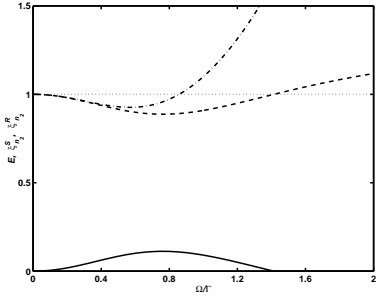,width=5cm}}
\end{center}
\caption{Entanglement measure $E$ (solid line) and the spin
squeezing parameters $\xi_{\vec{n}_{2}}^{S}$ (dashed line),  
$\xi_{\vec{n}_{2}}^{R}$ (dashed-dotted line) as a function of 
$\Omega/\Gamma$. }
\label{fig3}
\end{figure}

Thus, we conclude that the system is entangled when 
$|\rho_{eg}|>\rho_{ss}/2$, and again the entanglement is related to 
the Kitagawa and Ueda's spin squeezing parameter. Figure~\ref{fig3} 
shows $E$ and the squeezing parameters as a function of $\Omega$. An 
entanglement appears for $\Omega <\sqrt{2}\Gamma$ and, as predicted, 
corresponds to the spin squeezing predicted by the parameter 
$\xi_{\vec{n}_{2}}^{S}$. 

Finally, we turn to the third model in which the atoms are driven 
simultaneously by coherent and squeezed vacuum fields. In this case 
the density matrix has the same form as Eq.~(\ref{eq17}) but with the 
density matrix elements now dependent on $\Omega$ and the squeezing 
parameters $N$ and $M$. Hence, the condition for entanglement 
$|\rho_{eg}|>\rho_{ss}/2$ holds. However, it can be shown that one of 
the roots of Eq.~(\ref{eq19}) can be negative indicating that one can 
observe entanglement without spin squeezing. We have checked numerically
that this can happen for $\Omega \neq 0$ and $M>0$. For $M<0$ the 
roots are positive for all values of $\Omega$ and $N$. Hence, the 
condition for entanglement, $|\rho_{eg}|>\rho_{ss}/2$, also holds in 
this model and, according to Eq.~(\ref{eq6}), coincides with the 
condition for spin squeezing. 

\begin{figure}[h]
\begin{center}
\mbox{ \psfig{file=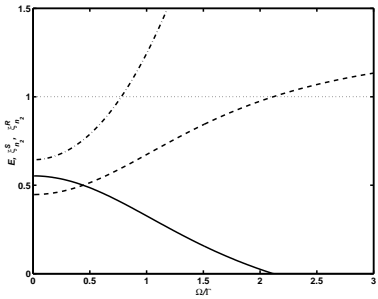,width=5cm}}
\end{center}
\caption{Entanglement measure $E$ (solid line) and the spin
squeezing parameters $\xi_{\vec{n}_{2}}^{S}$ (dashed line),  
$\xi_{\vec{n}_{2}}^{R}$ (dashed-dotted line) as a function of 
$\Omega/\Gamma$ for $N=0.1$ and $M=-\sqrt{N(N+1)}$.}
\label{fig4}
\end{figure}

We now present some numerical calculations that illustrate the above 
remarks. Figure~\ref{fig4} shows the entanglement measure and the 
squeezing parameters as a function of $\Omega$ for $N=0.1$ and 
$M=-\sqrt{N(N+1)}$. It is evident that similar to the models 
considered above, the entanglement is related to the spin squeezing 
given by $\xi_{\vec{n}_{2}}^{S}$. The entanglement induced at $\Omega =0$ 
decreases with increasing $\Omega$ and vanishes at 
$\Omega \approx 2.1\Gamma$, 
indicating that both entanglement and spin squeezing can be observed 
only for weak driving fields.

In summary, we have examined the relationship between entanglement and 
spin squeezing parameters in the two-atom Dicke system. 
Characterizing the spin squeezing parameters by the density matrix 
elements, we have examined simple models of driven two-atom Dicke 
systems in which spin squeezing and entanglement arise dynamically. 
We have found that the Kitagawa and Ueda's spin squeezing parameter is
a better measure of entanglement than the spectroscopic spin squeezing 
parameter. For the models discussed we have established that
the Kitagawa and Ueda's parameter is the sufficient and necessary 
condition for entanglement and the degree of entanglement is equal to 
the degree of spin squeezing.


\end{document}